\documentclass{IEEEtran}
\usepackage{cite}
\usepackage{amsmath,amssymb,amsfonts}
\usepackage{algpseudocode}
\usepackage{algorithm}
\usepackage{graphicx}
\usepackage{textcomp}
\usepackage{xcolor}
\usepackage{threeparttable}
\usepackage[T1]{fontenc}
\usepackage{gensymb}
\usepackage{booktabs, siunitx}
\usepackage{hyperref}
\def\BibTeX{{\rm B\kern-.05em{\sc i\kern-.025em b}\kern-.08em
    T\kern-.1667em\lower.7ex\hbox{E}\kern-.125emX}}

\ifCLASSOPTIONcompsoc
    \usepackage[caption=false, font=normalsize, labelfont=sf, textfont=sf]{subfig}
\else
\usepackage[caption=false, font=footnotesize]{subfig}
\fi

\renewcommand\vec{\mathbf}

\begin{document}
\title{Ray Launching-Based Computation of Exact Paths with Noisy Dense Point Clouds}

\author{\IEEEauthorblockN{Niklas Vaara,
Pekka Sangi, Miguel Bordallo L\'opez, and
Janne Heikkil\"a}\\ 
\IEEEauthorblockA{Center for Machine Vision and Signal Analysis\\
Faculty of Information Technology and Electrical Engineering, University of Oulu\\
Oulu, Finland\\
Email: niklas.vaara@oulu.fi}
\thanks{This work has been submitted to the IEEE for possible publication. Copyright may be transferred without notice, after which this version may no longer be accessible.}
}

\maketitle

\begin{abstract}
Point clouds have been a recent interest for ray tracing-based radio channel characterization, as sensors such as RGB-D cameras and laser scanners can be utilized to generate an accurate virtual copy of a physical environment. In this paper, a novel ray launching algorithm is presented, which operates directly on noisy point clouds acquired from sensor data. It produces coarse paths that are further refined to exact paths consisting of reflections and diffractions.
A commercial ray tracing tool is utilized as the baseline for validating the simulated paths. A significant majority of the baseline paths is found. The robustness to noise is examined by artificially applying noise along the normal vector of each point. It is observed that the proposed method is capable of adapting to noise and finds similar paths compared to the baseline path trajectories with noisy point clouds. This is prevalent especially if the normal vectors of the points are estimated accurately. Lastly, a simulation is performed with a reconstructed point cloud and compared against channel measurements and the baseline paths. The resulting paths demonstrate similarity with the baseline path trajectories and exhibit an analogous pattern to the aggregated impulse response extracted from the measurements. Code available at \url{https://github.com/nvaara/NimbusRT}.
\end{abstract}

\begin{IEEEkeywords}
Graphics processing unit (GPU), point cloud, radio propagation,  ray launching (RL), ray tracing (RT)
\end{IEEEkeywords}

\section{Introduction}

Ray tracing (RT) is a widely used approach for radio channel characterization. Especially for 6G, as we approach higher frequencies resulting in higher path losses, simulations will play a significant part in the design and deployment of network systems. Furthermore, high bandwidth is required for applications such as multi-user volumetric video streaming \cite{kamari2023environment}. Efficient and dynamic simulations for 6G integrated sensing and communications (ISAC) applications are the key enablers \cite{liu2023integrated}, which makes RT an attractive approach due to the significant speedups as a result of advances in  graphics processing unit (GPU) accelerated RT simulations \cite{yun2015ray}. In addition, ISAC requires physically accurate simulations of the environment, which is not possible with stochastic channel models \cite{hoydis2022sionna}.

Digital twins are envisioned to be the foundational technology behind 6G network simulation and planning \cite{lin20236g}. They encompass a physically accurate virtual model of the environment, which are constructed using multi-modal sensor data. They may be utilized to perform real-time RT simulations in the future \cite{alkhateeb2023real}. Furthermore, RT can be a useful tool for simulating how different sensors behave in the virtual environment. For example, in \cite{thieling2020scalable}, a RT-based approach was presented to produce realistic RADAR sensor simulations.

Many of the wireless communication components will be intelligent and driven by artificial intelligence and machine learning (ML) algorithms \cite{latva2019key}. For ML model training, RT can be used to generate highly accurate, site specific datasets \cite{alkhateeb2023real}. In \cite{moilanen2023ray}, data generated by ray tracing were utilized in conjunction with a radar image as an input to a convolutional neural network with the goal of detecting targets in a static environment. Moreover, RT can be utilized to study how these different components behave in an environment. An example of such component is reconfigurable intelligent surface (RIS), which is used to reflect signals to desired directions in order to improve the channel performance \cite{zhang2022active}. RT has been shown to be a great application for studying RIS channel characteristics \cite{pyhtilae2023ray, huang2022novel, huang2023ray, choi2023withray}.

RT-based methods for radio channel characterization are divided into the image method and ray launching (RL) algorithms \cite{yun2015ray}. Image method algorithms find the exact path trajectories in a given environment for a single transmitter (TX) and receiver (RX) pair. RL algorithms on the other hand cast rays from the TX in given directions without considering the locations of RXs. The rays interact with the scene geometry until a RX is hit or until the ray termination criteria is fulfilled. The latter is generally faster, especially when multiple RXs are considered, as the computations do not have to be performed for each RX separately. However, the produced paths are less accurate due to the discrete launching of rays from TXs, and as the RXs are often modeled as spheres to enable intersections. A variation of RL, known as environment-driven RL \cite{lu2018discrete} only casts rays to discretized points that were determined visible by visibility analysis. The visibility analysis is a pre-processing step which fills a matrix that describes the visibilities between all of the discretized elements in the environment. The visibility matrix enables quick intersection determination during the path finding process; however, the computational complexity of filling such matrix grows considerably as the number of discretized elements is increased.

Triangle mesh models are often utilized by RT-based algorithms, as triangles enable straightforward intersection determination due to their concise surface representation. As digital twin models will be formed from sensor data, the sensor-based 3D reconstruction will implicitly produce point clouds. Thus, utilizing the generated point clouds directly becomes an interesting application for RT-based methods.

Point clouds have been used for radio channel characterization by simulating various propagation effects \cite{jarvelainen2016indoor, jurveluinen2014sixty, koivumaki2023ray, virk2015simulating, koivumaki2018study, koivumaki2021impacts}. The simulations in the aforementioned papers, however, cannot be considered as fully RL-based. RL methods have been applied to models consisting of polygonal shapes that were acquired from point clouds \cite{pang2021gpu, wahl2022wip, suga2023indoor, kamari2023environment}. Recently, a point cloud-based RL method which utilizes a similar environment discretization-driven approach as \cite{lu2018discrete} was proposed. It directly operates on downsampled point clouds \cite{koivumaki2022point} and is capable of simulating multiple interactions efficiently on a consumer grade GPU. However, the visibility matrix generation is time consuming due to the high number of discretized elements and point-based geometry representation. Furthermore, the resulting paths are coarse approximations of the exact paths due to the errors imposed by the environment discretization. Another recent approach is measurement-based ray launcher (MBRL), where rays are launched in the directions where multipath components were observed in channel measurements \cite{de2023analysis}.

The previous point cloud-based works utilize laser scanned point clouds aside from \cite{suga2023indoor, kamari2023environment}, where RGB-D images were used to reconstruct the scene. While laser scanners provide superior accuracy, they are slow and expensive, which makes them unattractive for large scale deployment. On the other hand, depth sensors such as RGB-D cameras are cheap and already available in many consumer devices. In addition, RGB images can be used for material segmentation, which is essential for accurate RT-based simulations due to varying material properties. Furthermore, devices such as Microsoft Azure Kinect provide infrared images \cite{azurekinect}. Such images can be fused with RGB images to enhance material segmentation \cite{liang2022multimodal, reza2023multimodal}.

This paper is an extension of our previous work \cite{vaara2024ray}, which proposes a GPU-based RL method for point clouds. It supports the computation of exact paths consisting of reflections and diffractions with synthetic point clouds. In this paper, we provide greater details and improvements to the implementation and extend the simulator to work with noisy dense point clouds. Furthermore, we validate our approach with a point cloud reconstructed from RGB-D images and compare the computed paths with channel measurements, as well as with paths computed by a commercial ray tracer with a synthetic model of the reconstructed environment.

The paper is structured in the following way. In Section \ref{section_pipelineDesc} the simulator components are described. Section \ref{section_implDetails} provides additional details of the implementation. Section \ref{section_experiments} presents the performance and validation experiments. In Section \ref{section_discussion}, the findings are discussed. Lastly, in Section \ref{section_conclusion} the paper is concluded.  

\section{Overview of the Simulator}
\label{section_pipelineDesc}

\begin{figure*}[t]
  \centering
  \begin{minipage}[]{0.63\linewidth}
    \centering
    \includegraphics[width=\linewidth]{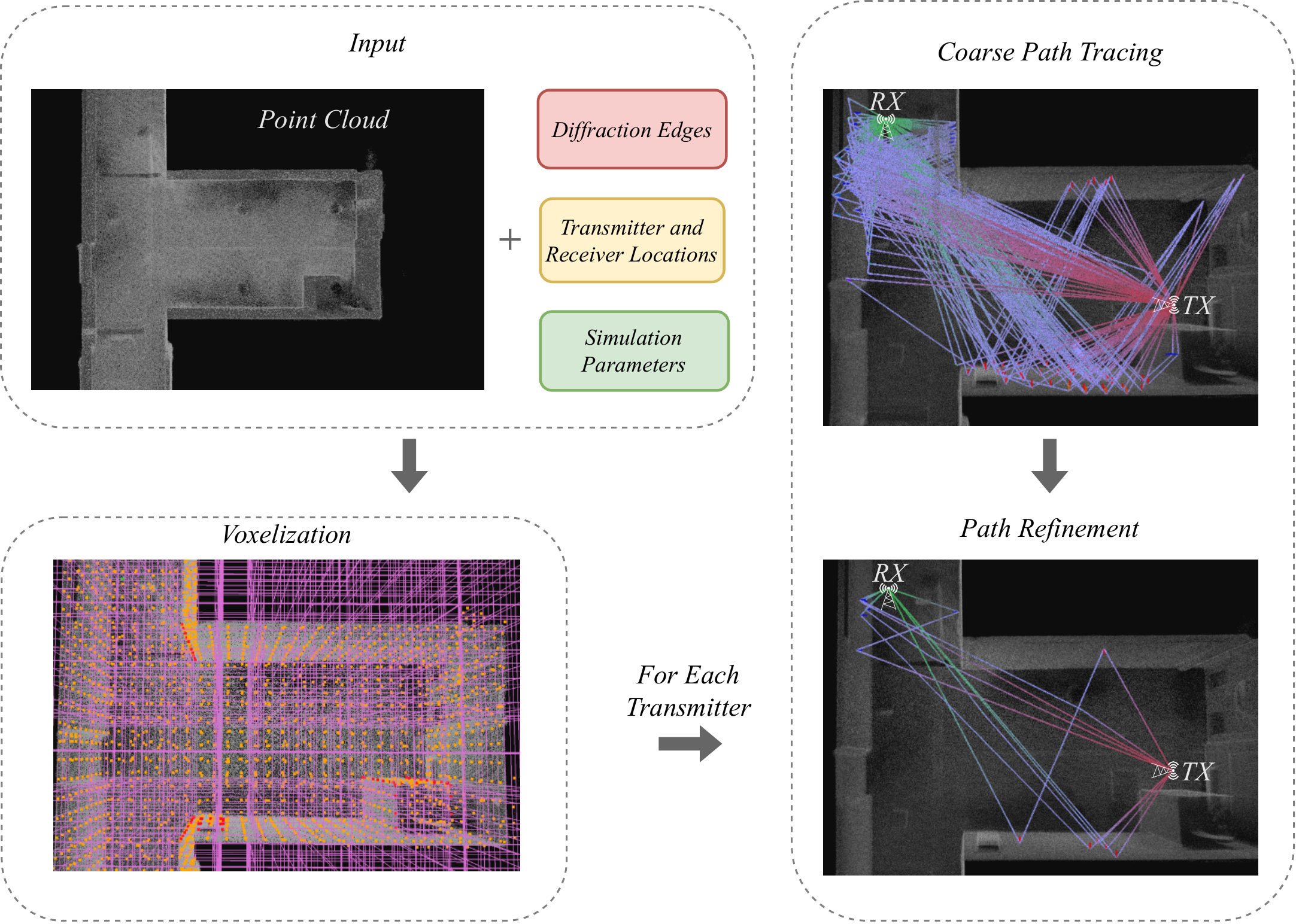}
    \caption{Illustration of the execution flow for a simulation consisting of two or fewer reflections. In voxelization, purple lines describe a voxel grid, while orange, red, and green points are the ray reception points used in the coarse path search.}
    \label{fig:pipeline_flow}
  \end{minipage}
  \hfill
  \begin{minipage}[]{0.335\linewidth}
    \centering
    \includegraphics[width=\linewidth]{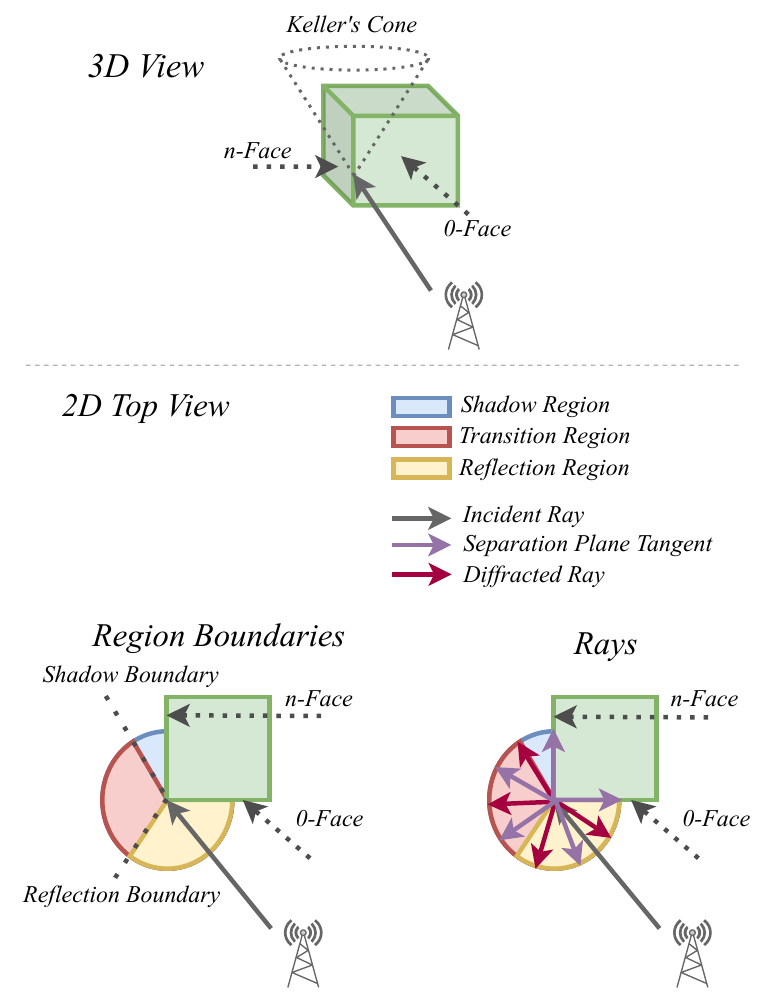}
    \caption{3D view of Keller's cone and 2D top view of its diffraction regions and rays.}
    \label{diffRegions}
  \end{minipage}
\end{figure*}

The proposed simulator outputs exact paths consisting of reflections and diffractions in a point cloud environment between a number of TXs and RXs. Main components of the implementation are voxelization, RL-based coarse path tracing, and path refinement, as shown in Fig.~\ref{fig:pipeline_flow}. In the following, they are presented with some details.
A shorter high-level description is provided in \cite{vaara2024ray}. 

\subsection{Input Details}

Each point in the point cloud contains a label, a normal vector, and a position. The labels are acquired by a segmentation process, where the resulting clusters represent distinct surfaces. Each point is then assigned the unique label of its respective cluster.

\label{ext_edge}
Considering diffraction, the point cloud is augmented with information about relevant edges. The simulator supports only exterior edges, which refer to edges that have an angle of $n \pi$, where $1 < n < 2$, between adjacent faces. The Uniform Theory of Diffraction (UTD) can also be applied to interior edges, where $(0 < n < 1)$  \cite{kouyoumjian1974uniform}. However, in the mm-wave band, diffraction has a non-negligible impact on the channel only near the incident shadow boundary (see Fig.~\ref{diffRegions}) in non-line-of-sight (NLOS) conditions \cite{jacob2012diffraction}. Since NLOS conditions are only present in the case of exterior edges, the simulator focuses on supporting exterior edge diffractions. This focus not only simplifies the model but also enhances the performance of the simulator.

\subsection{Voxelization}
\label{ss_voxelization}

A basic operation in ray tracing is calculation of intersections. To perform 
it efficiently with point clouds, we have to represent subsets of points as
volumetric objects. For this purpose, we utilize a voxelization process, where the environment is initially discretized with a given resolution
into small uniform elements called voxels.
Each voxel represents a part of the original environment bound by its volume. Due to uniformity, 
3D coordinates in the original environment can easily be transformed into voxel-space coordinates, and vice versa. The representation based on a voxel grid 
provides a coarse approximation of the underlying geometry, which is attractive for ray traversal algorithms due to simplicity and efficiency of spatial computations. Efficiency of such grid traversal has been observed, for example, in \cite{de2023analysis}. 

Generation of the voxel grid over the whole scene provides the highest level of the three-level voxelization process.
As voxel grids consume easily lots of memory, the resolution of this grid is kept as low as possible, while maintaining a representation of the environment where coarse characteristics are observable.
At the next level, we spatially divide each voxel into subvoxels with a given division factor $D_{v}$. 
Subvoxels are intermediate elements that are used for the sole purpose of forming intersectable entities (IEs). Each subvoxel may contain a set of points from the point cloud, diffraction edge segments, or RXs. The different types of IEs contain a label and a ray reception point, which are defined for each IE type in the following way:
\begin{enumerate}
    \item Point cloud primitive IEs (PCIE): The ray reception point equates to the average position of all point primitives within the subvoxel. The label is the label of the closest point to the IE voxel center point.
    \item Diffraction edge IEs (DEIE): The ray reception point is positioned at the edge segment's midpoint. Each edge has a unique label, which is shared by the IEs that are formed from the same edge.
    \item RX IEs (RXIE): The ray reception point corresponds directly to the RX's location. Each RX has a unique label.
\end{enumerate}
To counter the memory consumption of the high resolution voxel grid, each voxel is aware of its IEs. Thus, the IEs can be written linearly into memory. In addition, our implementation utilizes conical rays, which benefits from querying the lower resolution voxel grid for potential intersections. Lower resolution allows us to query broader parts of the voxel grid volume, after which we can utilize the IEs for more detailed intersection processing. Fig.~\ref{fig:pipeline_flow} contains an illustration of a low resolution voxel grid and IE ray reception points.

The lowest level of the voxelization process addresses
RT application programming interfaces (APIs) such as OptiX \cite{parker2010optix}, which utilize axis-aligned bounding boxes (AABBs) as the initial intersection primitives. We form these AABB primitives by dividing the subvoxels with a division factor $D_{sv}$. The reason for this spatial division is to minimize 
the number of points processed
during intersection determination. Similarly to the IE formation, only the AABBs that contain points are formed. Each AABB primitive is aware of the PCIE it belongs to, which allows us to determine the intersected PCIE during ray traversal.

The resulting AABB primitives may contain a point set, which only occupy a small portion of the AABB. This may cause unnecessary intersections, and thus, additional computational overhead. However, minimizing the AABB size based on the point positions is not optimal, as this may introduce a gap between neighboring AABB primitives. Such a gap may cause a ray to pass two neighboring AABB primitives that in reality represent a continuous surface. To minimize this effect, AABB primitives use their maximum length along a given axis when the maximum distance between two points within the AABB along that axis exceeds half of the maximum length.

\subsection{Intersections}

\label{intersection_tests}
\begin{figure}[htbp]
\centering
    \includegraphics[width=0.65\linewidth]{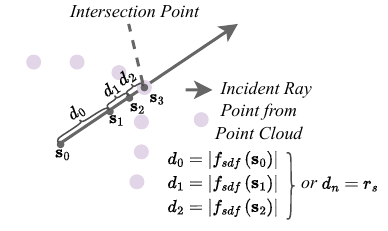}
    \caption{Illustration of the intersection determination procedure with a set of points. $r_{s}$ is used if $f_{sdf}$ fails.}
    \label{fig:sdf_march}
\end{figure}

When a ray is cast through a PCIE, it may intersect with some if its AABB primitives. Each hit of an AABB primitive involves determination of the intersection point
with a signed distance function (SDF)
\cite{adamson2004approximating}. 
Specifically, we estimate the distance to the surface with the implicit surface
equation
\begin{equation}
    \label{fSdf}
    f_{sdf}(\vec{x}) = \left(\vec{x} - \vec{p} \left(\vec{x}\right) \right) \cdot \vec{n}\left(\vec{x} \right), 
\end{equation}
where
$\vec{x}$ is a position being evaluated for intersection, $\vec{p}(\vec{x})$ and $\vec{n}(\vec{x})$ represent the weighted averages of the point positions and normals, respectively. The vectors $\vec{p}(\vec{x})$ and $\vec{n}(\vec{x})$ are evaluated in a similar manner \cite{adamson2004approximating}:
\begin{equation}
    \vec{p}(\vec{x}) = \frac{\sum_{i}^{} w\left(\| \vec{p}_{i} - \vec{x}\| \right) \vec{p}_{i}}{ \sum_{i}^{} w\left(\| \vec{p}_{i} - \vec{x}\| \right)}
\end{equation}
and
\begin{equation}
\label{nx}
    \vec{n}(\vec{x}) = \frac{\sum_{i}^{} w\left(\| \vec{p}_{i} - \vec{x}\| \right) \vec{n}_{i}}{ \sum_{i}^{} w\left(\| \vec{p}_{i} - \vec{x}\| \right)},
\end{equation}
where $\vec{p}_i$ and $\vec{n}_i$ are the $i$th position and normal vector of the set of points, respectively. Similarly to others \cite{adamson2004approximating, hubo2006quantized, kolluri2008provably}, for $w(x)$ we utilize the Gaussian weight function
\begin{equation}
\label{wx}
    w(x) = \exp\left({-\frac{x^2}{2 \sigma ^2}}\right)\text{.}
\end{equation}
We define $\sigma$ as $\xi r_{s}$, where $\xi$ and $r_{s}$ are user-defined scaling factor and sample radius, respectively. The sample radius $r_{s}$ is additionally used in the following ray marching process. A line segment is created with the length $l_{d} / (D_{v} D_{sv})$, where $l_{d}$ is the diameter of a voxel. The center point of the line segment is calculated by orthogonal projection of the AABB's center point to the ray. The initial sample position $\vec{s}_{0}$ is taken from the formed line segment end point closest to the ray origin. Starting from $\vec{s}_{0}$, we perform ray marching along the ray direction vector until an intersection is found or until the other end of the line segment is reached. The march distance is based on the absolute distance of the evaluated signed distance acquired with equation~(\ref{fSdf}), or $r_{s}$ if the calculation of signed distance fails. It may fail in situations where the distance between the $i$th sample position $\vec{s}_{i}$ and the points being evaluated is large, which leads to equation~(\ref{wx}) yielding a zero result. A ray-surface intersection is found if $ \text{sign} \left( f_{sdf} \left(\vec{s}_i \right)\right) \neq  \text{sign} \left( f_{sdf} \left( \vec{s}_{i+1} \right) \right)$ or if $|f_{sdf} \left(\vec{s}_i \right)|$ is below a user-defined threshold $t_{sdf}$. This procedure is visualized in Fig.~\ref{fig:sdf_march}.
In the case of RXIE and DEIEs, an intersection is found if the IEs are not occluded by PCIEs. By occlusion we mean that the rays cast towards RXIEs and DEIEs may intersect with PCIEs present in the scene.

\subsection{Coarse Path Tracing}
\label{coarse_path_tracing}

Coarse path tracing consist of transmission and propagation phases. Each of these phases is performed for each TX separately. As the name implies, the resulting paths are coarse approximations of the exact paths.

\subsubsection{Transmission Phase}
In traditional RL algorithms, rays are initially launched from the TX with uniform spacing, as noted in \cite{yun2015ray}. This method often encounters an issue due to the increasing spatial separation between neighboring diffracted rays. Although this problem can be alleviated, it often requires extra computational effort. One commonly used technique to address this is ray splitting. This involves creating new rays to ensure that neighboring rays maintain a consistent spatial distance from one another, as described in \cite{fortune1998efficient}.
However, our approach diverges from this method. Taking inspiration from the environment-driven approaches in \cite{lu2018discrete, koivumaki2022point}, we launch initial rays from the TX directed at all IE reception points in the environment. This phase produces the initial interactions with visible IEs from the TX and establishes line of sight (LOS) connections to RXs.

\subsubsection{Propagation Phase}
In this phase, coarse paths consisting of reflections and diffractions that reach a RX are discovered. PCIE and DEIE interactions determined by the transmission phase provide starting points for the propagation phase. The rays are traced until the termination criteria is fulfilled. The ray is terminated if it does not intersect with any IEs. In addition, we define an upper bound for the number of interactions and diffractions. Instead of time consuming pre-processing to generate a visibility matrix as in \cite{koivumaki2022point}, we utilize the voxel grid presented in Section \ref{ss_voxelization} for traversal. Together with conical rays, we refer this method of traversal as voxel cone tracing.

\subsubsection{Voxel Cone Tracing}
\label{voxel_cone_tracing}
\begin{figure}[htbp]
    \centering
    \includegraphics[width=0.7\linewidth]{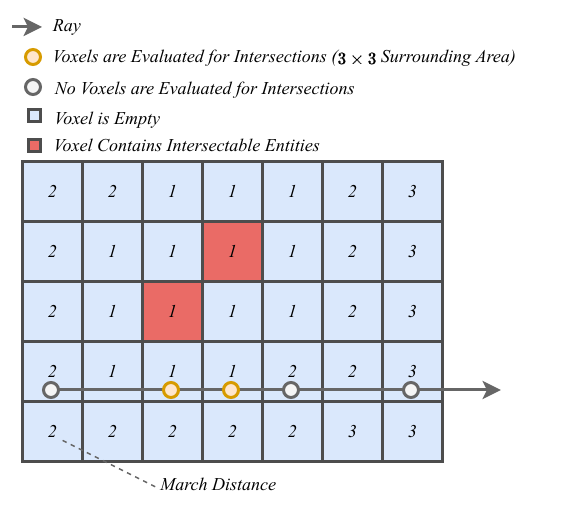}
    \caption{2D illustration of voxel ray marching steps. The number inside each voxel is the march distance constrained to a minimum of one.}
    \label{fig:traversal_illustration}
\end{figure}

Our voxel grid traversal algorithm utilizes ray marching \cite{pharr2005gpu}, which refers to advancing a varying distance during each iteration. The march distance from a given voxel is the longest distance along any axis to the closest voxel that contains IEs. Each voxel includes the information whether it contains IEs, as well as the march distance to the closest voxel which encompasses IEs. In such a case where the voxel itself contains IEs or if the march distance is equal to one, all of the neighboring voxels are evaluated for intersections. During each ray march iteration, a voxel history buffer is filled and compared against the previous sampled voxels to avoid duplicate processing. Voxel traversal is illustrated in Fig.~\ref{fig:traversal_illustration}.

\begin{figure}
  \centering
  \includegraphics[width=0.7\linewidth]{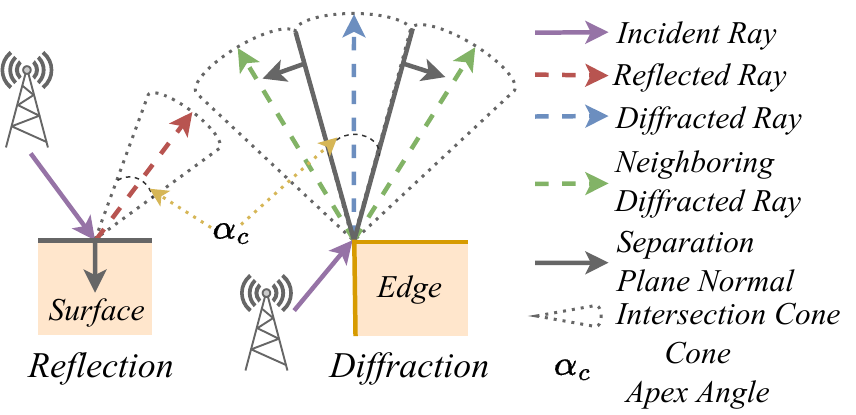}
  \caption{Attributes of conical reflection and diffraction rays used in the intersection tests. Note that in the case of diffraction multiple of these rays are cast from the edge in the shape of Keller's cone (Fig.~\ref{diffRegions}).}
  \label{refldiff}
\end{figure}

Similarly to \cite{koivumaki2022point, de2023analysis}, we utilize conical rays (see Fig.~\ref{refldiff}). They are used to determine intersections with IEs, as they result in a realistic approximation for the potential ray paths from a surface. Diffracted rays are often simulated by casting rays in the form of a Keller's cone \cite{keller1962geometrical}, see Fig.~\ref{diffRegions} for an illustration. This introduces increasing spatial separation between neighboring diffracted rays, which we account for by utilizing conical rays. The intersection cone apex angle $\alpha_{c}$ is determined based on the used voxel size and scene dimensions. It is chosen so that the cone radius is encompassed by the neighboring voxels at a distance equal to the diameter of the scene's AABB. Each conical diffraction ray is restricted by two neighboring separation planes, as illustrated in Fig.~\ref{refldiff}. A 2D illustration of diffracted rays and separation plane tangent vectors is shown in Fig.~\ref{diffRegions}.

For each voxel containing IEs that is being evaluated for intersections, a bounding sphere is formed around it for efficient cone-sphere intersection. If the sphere is inside the cone, the IEs inside the voxel are processed and a similar cone-sphere intersection test for each IE is performed, however, with a smaller bounding sphere that only encompasses the subvoxel. The exception to the rule is RXIEs when the number of interactions is less than or equal to two. In such a case the intersection is considered valid if the initial voxel-cone intersection in which the RXIE lies in is deemed valid. 
This is performed as the volume covered by a conical ray from the first intersection point is initially  small but grows as the number of interactions increases.
For example, the first conical ray may intersect with $N$ PCIEs, which results in $N$ new conical rays to be traced originating from a single source point. This way, we can reduce the potential loss of important paths caused by the discrete low resolution representation of IEs. If the IE cone-sphere intersection is valid, the visibility and intersection is validated by tracing a ray from the source point to the IE reception point. 

Due to the cone-sphere intersections, we use two separation planes for diffractions to mitigate duplicate intersections with the neighboring diffracted rays. We calculate the signed distance from the separation planes to the IE reception point in order to determine if it lies between the planes. In the case of reflected rays, two identical separation planes are placed at the origin of the reflection source. This is beneficial, as it removes the divergence of computations between the intersection testing of reflected and diffracted rays. The components of the aforementioned concial rays are illustrated in Fig.~\ref{refldiff}.

\subsubsection{Valid Intersection Processing}

The different IE types behave differently when an intersection is found. When a RXIE is encountered, the path is saved for further processing. However, as the coarse paths that reached a RX in many cases result in multiple similar paths due to the conical rays, only $\kappa$ shortest paths with a given label and interaction type combination are saved to speedup further processing. See Fig.~\ref{fig:pipeline_flow} for an example of coarse paths between a TX and RX. In the case of an intersection with a PCIE, a single ray is reflected. In the case of diffractions, multiple new rays with a given angular discretization restricted to the edge's exterior angle in the form of Keller's cone \cite{keller1962geometrical} are launched, as illustrated in Fig.~\ref{diffRegions}.

\subsection{Path Refinement}

The produced coarse paths represent approximate reflections and diffractions due to the conical rays and the additional error caused by discretized interaction points. A common method for correcting the paths found by RL is to apply the image method as a post-processing step to these paths, as in \cite{tan1996microcellular, hoydis2023sionna, choi2023withray, wirelessinsite}. However, image method cannot be directly used to find the exact diffracted ray direction, which in \cite{funkhouser2004beam} was addressed by solving systems of nonlinear equations. Recently, minimization algorithms that support reflections and diffractions for finding the exact paths have been developed \cite{vaara2023refined, eertmans2023min}. 

We utilize a similar approach as in our previous work \cite{vaara2023refined}, where the exact paths are computed with a gradient descent-based path length minimization scheme derived from the Fermat's principle of least time. The previous work assumes infinite planes for reflection points, which we addressed in \cite{vaara2024ray} and extend in this work by adapting the implementation to work with noisy point clouds. During the refinement, we dynamically update the normal vectors of reflections based on the determined intersection point. Thus, with this change, the algorithm adapts to the underlying geometry, which is relevant, as our implementation does not impose any simplifications to the geometry. However, the mathematical foundation of the algorithm is the same, which is the following. The function to be minimized is
\begin{equation}
    f = \sum_{k=0}^{N+1} ||\vec{I}_{k} - \vec{I}_{k+1}||,
\end{equation}
where $\vec{I}_{k}$ is the $k$th interaction point. In the case of $N$ interactions, $\vec{I}_{0}$ and $\vec{I}_{N+1}$ are the TX and RX locations, respectively. Thus, for each interaction point $\vec{I}_{k}, k \in [1, N]$, the local minimization problem is defined as
\begin{equation}
\label{f_k}
    f_{k}\left(\vec{I}_k\right) = ||\vec{I}_{k} - \vec{I}_{k - 1}|| + ||\vec{I}_{k} - \vec{I}_{k + 1}||\text{.}
\end{equation}
In the case of reflections, the $k$th reflection point along a path is defined as
\begin{equation}
    \label{refl_eq}
    \vec{I}_{k} = \vec{R}_{k} = \vec{c}_{k} + r_{k} \vec{u}_{k} + s_{k} \vec{v}_{k},
\end{equation}
where $\vec{c}_{k}$ is the current position determined by tracing a ray from $\vec{I}_{k-1}$ to $\vec{I}_{k}$, and $\vec{u}_{k}$ and $\vec{v}_{k}$ are vectors that form an orthonormal basis with the surface normal at $\vec{c}_{k}$. Lastly, $r_{k}$ and $s_{k}$ are the unknown variables that we aim to solve.

Diffraction points are defined as line segments, and thus, can only move along the edge. Similarly to reflections, the $k$th diffraction point is defined as
\begin{equation}
\label{diff_eq}
    \vec{I}_{k} = \vec{D}_{k} = \vec{c}_{k} + t_{k} \vec{w}_{k},
\end{equation}
where $\vec{c}_{k}$ is the current position on the line segment, $\vec{w}_{k}$ is the direction vector along the edge, and $t_{k}$ is the unknown variable to be solved.

During each iteration, for reflections and diffractions the gradient vector $\nabla f_{k}$ is formed by differentiating $f_{k}$ with respect to $r_{k}, s_{k}$ and $t_{k}$, which yields
\begin{align}
    \label{fkr}
    \frac{\partial f_{k}}{\partial r_{k}} = \left(\frac{(\vec{I}_{k} - \vec{I}_{k + 1})}{\|\vec{I}_{k} - \vec{I}_{k + 1} \| } + \frac{(\vec{I}_{k} - \vec{I}_{k - 1})}{\|\vec{I}_{k} - \vec{I}_{k - 1} \| }\right) \cdot \vec{u}_{k}, \\
    \label{fks}
    \frac{\partial f_{k}}{\partial s_{k}} = \left(\frac{(\vec{I}_{k} - \vec{I}_{k + 1})}{\|\vec{I}_{k} - \vec{I}_{k + 1} \| } + \frac{(\vec{I}_{k} - \vec{I}_{k - 1})}{\|\vec{I}_{k} - \vec{I}_{k - 1} \| }\right) \cdot \vec{v}_{k},
\end{align}
for reflections, and
\begin{equation}
    \label{fkt}
    \frac{\partial f_{k}}{\partial t_{k}} = \left(\frac{(\vec{I}_{k} - \vec{I}_{k + 1})}{\|\vec{I}_{k} - \vec{I}_{k + 1} \| } + \frac{(\vec{I}_{k} - \vec{I}_{k - 1})}{\|\vec{I}_{k} - \vec{I}_{k - 1} \| }\right) \cdot \vec{w}_{k},
\end{equation}
for diffractions. Each interaction point position is then updated by gradient descent. As in \cite{vaara2023refined}, we utilize backtracking line search for determining the step size $\gamma$ used in gradient descent. During each iteration starting from $\gamma = 1$, the step size is acquired when the following condition is fulfilled \cite{boyd2004convex}:
\begin{equation}
    f_{k}(\vec{x}_{k} + \gamma \Delta \vec{x}_{k}) > f_{k}(\vec{x}_k) + \alpha \gamma \nabla f_{k}^T \Delta \vec{x}_{k},
\end{equation}
where $\vec{x}_k$ is the current position, $\nabla f_{k}$ are the gradients, and $\Delta \vec{x}_{k} = -\nabla f_{k}$. The step size $\gamma$ is updated iteratively by multiplying it with $\beta$, which along with $\alpha$, is a user-defined parameter.

During each iteration the reflection points are determined by tracing a ray from $\vec{I}_{k-1}$ to the updated position evaluated by gradient descent. As reflection points are susceptible to noise, we extend our previous work \cite{vaara2024ray} to account for it in the following way. We aim to improve the convergence by introducing an additional refinement step to improve the quality of the normal vector. This is achieved by recalculating the normal vector with equation~(\ref{nx}) by additionally evaluating the points in the neighboring AABB primitives. The refined normal vector is then used to calculate the vectors $\vec{u}_{k}$ and $\vec{v}_{k}$ for the next iteration. However, the noisiness still imposes a problem for convergence. To further improve the convergence, we introduce distance and angle thresholds. 
The distance threshold $t_{d}$ defines the maximum distance from the new position to a plane formed at the previous position with the previous normal vector. The angle threshold $t_{a}$ is the maximum angle between the previous and new normal vectors. If the determined distance and angle are below the thresholds, the vectors $\vec{u}_{k}$ and $\vec{v}_{k}$ are not updated for the next iteration.
In the case of diffractions, the determined position by gradient descent does not require any additional processing, as $\vec{w}$ in equation~(\ref{diff_eq}) is constant for each edge.

Reflection intersection point $\vec{I}_{k}$ is acquired by tracing a ray from $\vec{I}_{k-1}$ if a surface is hit. In the case of diffractions, the new position is evaluated whether it still lies on the diffraction edge. If one of these steps fail, the refinement process is terminated for the path.
The refinement is performed iteratively until the maximum number of iterations $\rho$ is reached or the aforementioned validation steps during refinement fail. In the former case, the path is considered valid if the squared norm of the gradient vector consisting of reflections and diffractions $|| \nabla f ||^2$ is less than $\delta$, which is the user-defined convergence threshold for refined paths. Lastly, for each path that passed the refinement phase, the visibility from each interaction point to the previous and next interaction point is validated by RT.

Similarly to the coarse path tracing, the converged refined paths are processed to remove the duplicate paths. However, unlike in coarse path tracing, only the the shortest paths with a certain label and interaction combination are kept, as the refined paths represent the exact paths. In the case of reflection points, we determine a more accurate label by finding the label of the closest point to the intersection point. Lastly, as a post-processing step we calculate the first Fresnel zones for each interaction point for each path to further remove duplicate paths. This is beneficial, as some reflection points may converge near an edge, where the points contained in an IE consist of points from the edge faces that share different labels.

\begin{figure}[htbp]
    \centering
    \includegraphics[width=0.7\linewidth]{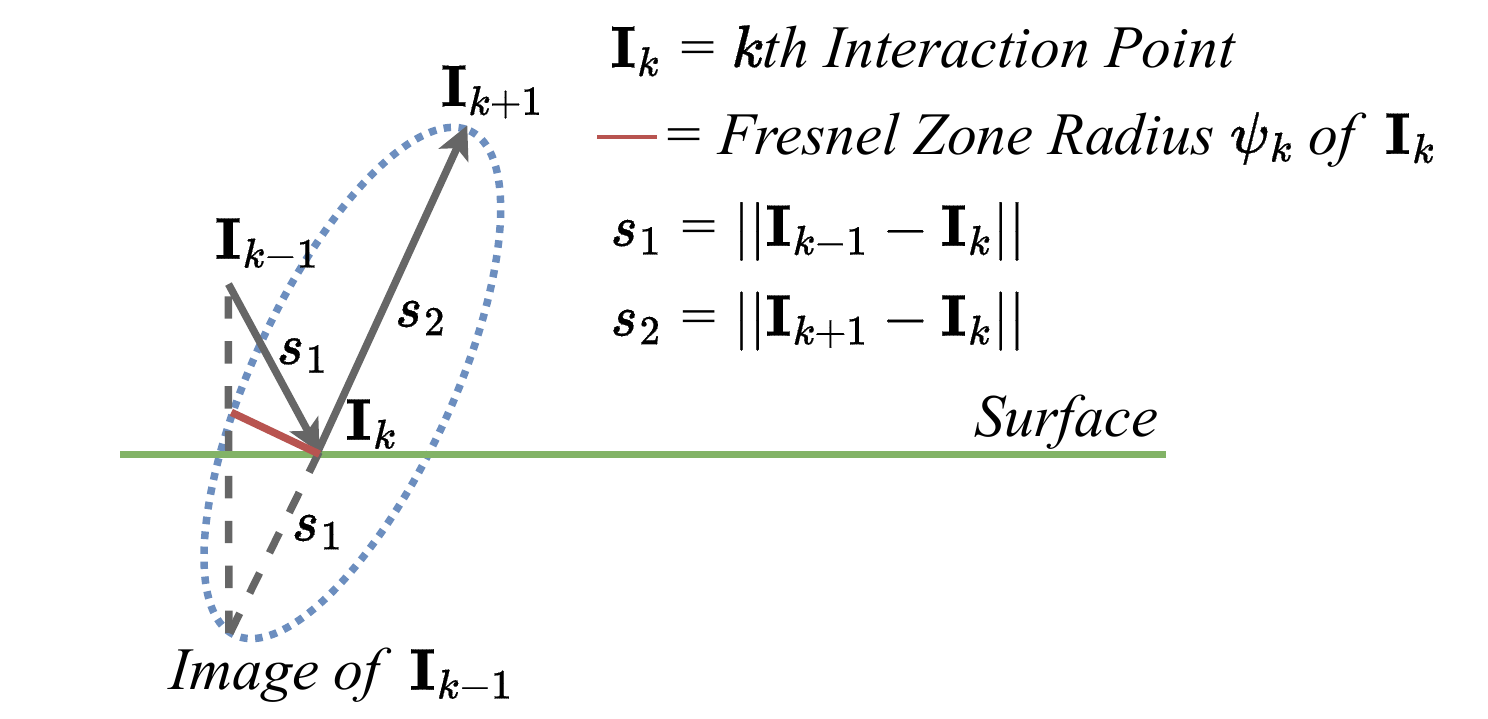}
    \caption{Illustration of the first Fresnel zone.}
    \label{FresnelZone}
\end{figure}

We utilize a similar method to \cite{jarvelainen2016indoor, koivumaki2022point}, where the first Fresnel zones are used to validate uniqueness of each path. In our implementation, we compare whether the interaction points of a specific path lie in the first Fresnel zone radii of another path. The radius $\psi_{k}$ at the $k$th interaction point $\vec{I}_{k}$ is calculated with
\begin{equation}
    \psi_{k} = \sqrt{\frac{\lambda s_{1}s_{2}}{s_{1} + s_{2}}},
\end{equation}
where $s_{1} = || \vec{I}_{k-1} - \vec{I}_{k}||$, $s_{2} = ||\vec{I}_{k+1} - \vec{I}_{k}||$, and $\lambda$ is the wavelength. Such scenario is illustrated in Fig \ref{FresnelZone}.

First, the refined paths are sorted based on propagation time delay. Starting from the lowest time delay, the uniqueness of the path is validated by comparing to the existing paths that were deemed valid. The path is considered to be a duplicate of an existing path if the interaction points are inside the first Fresnel zones of the existing path and if the chain of interaction types of both paths is equal. In addition, the trajectories of the paths must be similar, \textit{i.e.} the angle between the $k$th ray of both paths must be less than a given threshold. See Fig.~\ref{fig:pipeline_flow} for an example of resulting paths after the refinement and post-processing steps.

\section{Implementation Details}
\label{section_implDetails}
The simulator aside from post-processing and other minor parts, is GPU-based. The implementation is written in C++17. For GPU acceleration we use CUDA, and OptiX as the RT API.

\subsection{GPU Accelerated Ray Tracing}

\begin{figure}[htbp]
    \centering
    \includegraphics[width=0.5\linewidth]{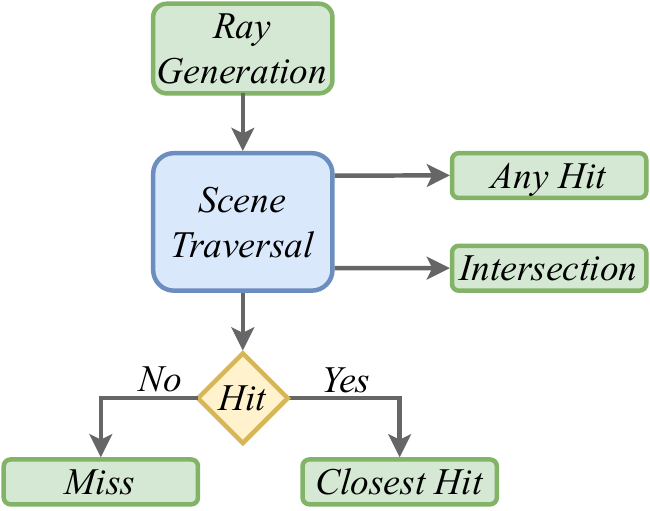}
    \caption{Ray tracing execution graph. User-defineable programs are denoted by green color.}
    \label{rtInterface}
\end{figure}

OptiX and other modern RT APIs such as Vulkan \cite{vulkan} work in a similar fashion. They provide an interface for intersection determination and processing using user-defined programs, which are illustrated in Fig.~\ref{rtInterface} by green color.

Ray generation programs are the entrypoint of each ray, from which the scene traversal process is started. The API-defined scene traversal process utilizes a spatial data structure known as acceleration structure to determine intersections with AABBs. These AABBs, as well as the intersection program are implicitly defined for triangle-based meshes by the RT API, as triangles are the most common primitives used in RT applications. For point clouds, a custom intersection program and AABBs have to be defined explicitly.

When an intersection program reports an intersection, the optionally defined any-hit program is called. After the scene traversal is finished, the closest hit program is called if an intersection was found. Otherwise, the miss program is invoked. Modern RT APIs also support other user-defineable programs, such as callables, which can be used to improve the performance of divergent code. The final output of these programs is determined by the information written into a user-defined payload associated with each ray.

\subsection{Pre-Processing} 

Pre-processing steps are critical in optimizing the traversal and intersection processes of RT, ensuring efficient and accurate results. The pre-processing measures include scene voxelization, acceleration structure building, and optimization of rays diffracted from the diffraction edges.

\subsubsection{Scene}

\label{sss_scene}

All of the steps involved in voxelization explained in Section \ref{ss_voxelization} are performed on the GPU. This can be done efficiently, as each voxel can be processed simultaneously. The resulting PCIE AABB primitives are given as a build input to the RT acceleration structure.
\label{3dTexFill}
For each voxel, an index that leads to the voxel's IEs, as well as the the longest distance along an axis to the closest voxel containing IEs (constrained to a minimum of one) are computed on the GPU. This information is then written into a 3D texture, which offers a texture cache and hardware-based out-of-bounds validation for efficient sampling \cite{nvidiatexture}.

\subsubsection{Diffraction Edges}
The combined area of shadow, transition, and reflection region illustrated in the 2D top view  of Fig.~\ref{diffRegions} showcases the area of diffracted rays on the 2D plane orthogonal to the edge direction, which are calculated as a pre-processing step for each edge separately as shown in the rightmost illustration of Fig.~\ref{diffRegions}. As we use conical rays, it enables for optimizations. For example, if each 2D diffraction ray has an apex angle of $\alpha_{c}$. When the rays are transformed into 3D space, the angle between these rays decreases as the angle between the edge direction vector and reflected ray from 0-face is decreased. Thus, we calculate the optimal number of diffraction rays for a specific angle with a given discretization resolution. 

\begin{figure}[htbp]
    \centering
    \includegraphics[width=0.5\linewidth]{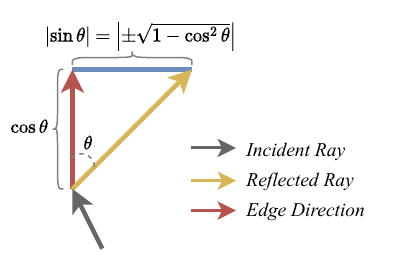}
    \caption{Variables based on the reflection angle which are utilized to compute the 3D transformation of the 2D diffraction vectors in (\ref{diff_transform}).}
    \label{diffDistances}
\end{figure}

The process of transforming the rays to 3D space is the following. Each pre-processed diffraction ray has a direction and two separation plane tangent vectors, as illustrated in Fig.~\ref{diffRegions}. The latter vectors define the bounds for each conical diffraction ray. The edge orientation can be defined with a rotation matrix $\vec{R}$, where the forward direction is $\vec{e}$. With $\vec{R}$ and the reflected ray direction, the 3D transformation of the pre-computed diffraction rays can be expressed by
\begin{equation}
\label{diff_transform}
    \vec{x}_{3D}(\vec{x}) = \vec{R} \vec{x} |\sin(\theta)| + \vec{e} \cos(\theta),
\end{equation}
where $\vec{x}_{3D}$ is the resulting 3D vector transformed from the input vector $\vec{x}$ which lies on the 2D plane orthogonal to $\vec{e}$ in the local coordinate system of the edge, and $\theta$ is the angle between edge forward direction $\vec{e}$ and reflected ray direction, illustrated in Fig.~\ref{diffDistances}. Lastly, the 3D separation plane normal vectors are computed with a normalized cross product of the 3D separation plane directions and $\pm \vec{e}$, where the sign of $\vec{e}$ determines the separation plane normal vector direction. In our implementation the direction is outwards from the diffraction ray direction, as illustrated in Fig.~\ref{refldiff}.

\subsection{Intersections and Traversal}
The intersection testing and traversal are performed on the GPU, as such operations are computed for each ray independently. In addition, modern GPUs contain hardware accelerated ray tracing cores, which boost the performance of intersection testing and ray traversal operations. Our implementation can only exploit the latter enhancement.

\subsubsection{Intersections and Visibility Validation}
Each ray-based visibility validation or intersection point determination is performed using OptiX. Diffraction edge and RXIEs are not considered to be part of the acceleration structure geometry, as they are only used as the target point for visibility validation. The aforementioned IE types are considered to be visible if the miss program is invoked. As diffraction edges lie on top of PCIEs, we reduce the maximum length of the cast ray by a small bias to avoid such intersection. PCIEs form the intersectable geometry, as they are used to build the acceleration structure. Since we utilize point clouds, a custom intersection program has to be defined. Our intersection program evaluates the intersection point and normal vector as explained in Section \ref{intersection_tests}.

\subsubsection{Voxel Traversal}

As mentioned in Section \ref{3dTexFill}, the 3D texture represents the voxel grid in texture memory. The traversal process is performed by ray marching in the ray direction based on the march distance acquired by sampling the texture. A value can be defined in the texture sampler which is returned when an out-of-bounds sampling occurs. We utilize such value as a signal for traversal termination.

Voxel-based ray traversal algorithms optimized for GPU computing have been developed, such as \cite{xiao2012efficient}. We further developed a voxel ray marching algorithm which utilizes the distance information provided by the texture. The operations of the algorithm are demonstrated in Alg. \ref{alg:ray_march}. For additional clarification, we provide details on the introduced variables. The algorithm input consists of the following variables: Voxel space position $\vec{V}_{pos}$, ray direction $\vec{R}_{dir}$, and $a_{dist}$, which is the march distance to a voxel that contains IEs. The march distance is acquired by sampling the texture at $\vec{V}_{pos}$. It calculates the minimum step size to travel $a_{dist}$ along any axis and returns the new voxel space position $\vec{V}_{new}$. By steps we refer to the factor $S_{total}$ in $\vec{V}_{new} = \vec{V}_{pos} + \vec{R}_{dir} S_{total}$. The algorithm first calculates the voxel origin $\vec{C}$ and $\vec{L}$, where each element represents positivity of the sign of the corresponding element in $\vec{R}_{dir}$. $\vec{S}_{unit}$ is the inverse of $\vec{R}_{dir}$ where each element is constrained to a minimum value of $\epsilon_{z}$ to avoid division by zero. Each of its elements represents the distance needed to travel in the direction of $\vec{R}_{dir}$ in order to move one unit along the respective axis. $\vec{D}_{next}$ is the distance from $\vec{V}_{pos}$ to the next voxel along each axis with respect to $\vec{R}_{dir}$. The steps to the next voxel $\vec{S}_{next}$ along each axis is acquired by the Hadamard product of $\vec{D}_{next}$ and $\vec{S}_{unit}$. The total number of steps $\vec{T}$ along each axis can then be determined by addition of $\vec{S}_{next}$ and $\vec{S}_{unit}(N-1)$. Lastly, the minimum value of the components of $\vec{T}$ is evaluated by exploiting AND operations to form $\vec{X}$ where only the smallest component is equal to one. The total step size $S_{total}$ can then be retrieved by a dot product of $\vec{T}$ and $\vec{X}$. In addition, a small bias $\epsilon_{s}$ is added to the total step size to account for floating point errors. 

\algrenewcommand\algorithmicrequire{\textbf{Input:}}
\algrenewcommand\algorithmicensure{\textbf{Output:}}

\begin{algorithm}
\caption{Voxel Ray Marching Algorithm}\label{alg:ray_march}
\begin{algorithmic}
\Require{$\vec{V}_{pos}, \vec{R}_{dir}, a_{dist}$} \Comment{ $a_{dist} \geq 1$}
\Ensure{$\vec{V}_{new}$}
\State $\epsilon_{s} = 1\text{e-}2$ \Comment{Step size error bias}
\State $\epsilon_{z} = 1\text{e-}16$ \Comment{Zero division bias}
\State $\vec{C} = \text{floor}(\vec{V}_{pos})$
\State $\vec{L} = \vec{R}_{dir} \geq 0$ \Comment{Element-wise comparison}
\State $\vec{S}_{unit} = 1 / \text{max}(\text{abs}(\vec{R}_{dir}), \epsilon_{z})$ \Comment{Steps per unit}
\State $\vec{D}_{next} = \text{abs}(\vec{L} - (\vec{V}_{pos} - \vec{C}))$ \Comment{Distance to next voxel}
\State $\vec{S}_{next} = \vec{D}_{next} \odot \vec{S}_{unit}$ \Comment{Steps to next voxel}*
\State $\vec{T} = \vec{S}_{next} + \vec{S}_{unit}(N - 1)$ \Comment{Total steps along each axis}
\State $\vec{X}_{x} = (\vec{T}.x \leq \vec{T}.y) \And (\vec{T}.x \leq \vec{T}.z)$
\State $\vec{X}_{y} = (\vec{T}.y <  \vec{T}.x) \And (\vec{T}.y \leq \vec{T}.z)$
\State $\vec{X}_{z} = (\vec{T}.z <  \vec{T}.x) \And (\vec{T}.z <  \vec{T}.y)$
\State $\vec{X} = [\vec{X}_{x}, \vec{X}_{y}, \vec{X}_{z}]^T$ \Comment{For deciding the step size}
\State $S_{total} = \vec{T} \cdot \vec{X} + \epsilon_{s}$ \Comment{The final step size}
\State $\vec{V}_{new} = \vec{V}_{pos} + \vec{R}_{dir}S_{total}$
\State return $\vec{V}_{new}$
\end{algorithmic}
\begin{threeparttable}
\begin{tablenotes}
\item[*] Operator $\odot$ is the Hadamard product.
\end{tablenotes}
\end{threeparttable}
\end{algorithm}

\section{Experiments}
\label{section_experiments}

In this section, we first evaluate the performance and validate our simulation results against Wireless Insite \cite{wirelessinsite}. We then assess how the introduction of noise and normal vector estimation affect the results using a synthetic model. As reconstructed point clouds are the intended geometry to be used in our simulator, we further validate our approach with such data. In addition, it is important to see how well the simulated paths correlate with actual channel measurements. Thus, the simulated paths, as well as paths acquired with Wireless Insite are compared to an aggregated impulse response extracted from channel measurements.

\subsection{Scene Description}

\begin{figure*}[t]
\centering
    \includegraphics[width=1.0\linewidth]{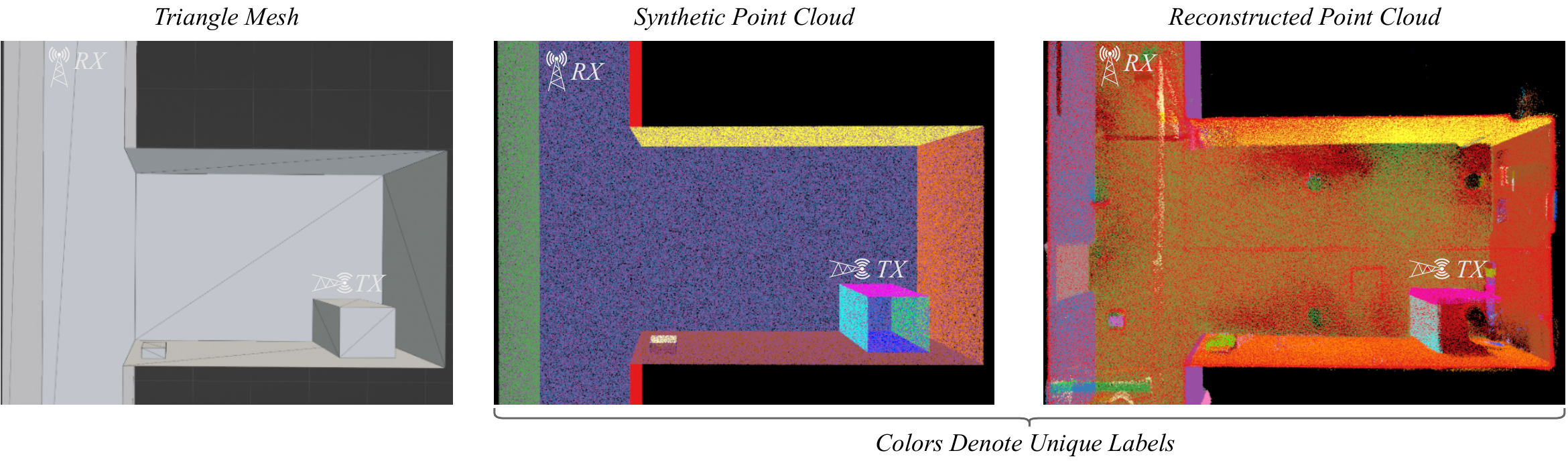}
    \caption{Models used in the experiments. The ceiling is hidden in the triangle mesh model for visualization purposes.}
    \label{fig:models}
\end{figure*}

The synthetic model used in the experiments is a triangle mesh model of a corridor. Such model is needed for the experiments, as Wireless Insite does not support point clouds as the geometry input. For our proposed implementation, the aforementioned model was turned into a point cloud containing a total of 2~745~513 points. In addition, the corridor was reconstructed using BS3D \cite{mustaniemi2023bs3d} with Microsoft Azure Kinect depth camera, which resulted in a point cloud consisting of 2~727~608 points. The labels for both of these point clouds were generated with a RANSAC-based plane fitting algorithm available in Point Cloud Library (PCL) \cite{rusu20113d}. The triangle mesh, as well as the point clouds are shown in Fig.~\ref{fig:models}.

\subsubsection{Simulation Parameters}

The simulations use different parameters depending on the input point cloud data. Generic parameters which are the same across all of the simulations are presented in Table \ref{tab:generic_params}. Simulation specific RT parameters utilized by coarse path tracing and path refinement are presented in Table \ref{tab:rt_params} and Table \ref{tab:refine_params}, respectively. "Noisy" and "Noiseless" in the aforementioned tables refer to the presence or lack of noise in the input point cloud. In the scenarios where estimated normal vectors are used, they were approximated by considering the neighboring points of each individual point in the radius of 10cm by a least squares plane fitting algorithm available in PCL.

\begin{table}[htbp]
    \centering
    \caption{Generic Parameters}
    \begin{tabular}{cc}
        \hline
        Parameter & Value\\
        \hline
        Carrier Frequency $f$ (GHz) & 60 \\
        Voxel Division Factor $D_{v}$ & 2\\
        Subvoxel Division Factor $D_{sv}$ & 4\\
        Scaling Factor $\xi$ & 2.0 \\
        Coarse Path Duplicate Limit $\kappa$ & 100 \\
        Refinement Iterations $\rho$ & 2000 \\
        Convergence Threshold $\delta$ & 0.0001\\
        Line Search Parameter $\alpha$ & 0.4\\
        Line Search Parameter $\beta$ & 0.4\\
        \hline
    \end{tabular}
    \label{tab:generic_params}
\end{table}
\begin{table}[htbp]
    \centering
    \caption{Ray Tracing Parameters}
    \begin{tabular}{SSSSSS[table-format=3.0]}
         \hline
         {Case} & {$r_{s}$ (m)} & {$t_{sdf}$(m)}\\
         \hline
         {Coarse Path Tracing} & 0.015 & 0.0015\\
         {Noiseless, Refinement} & 0.003 & 0.0005\\
         {Noisy, Refinement} & 0.01 & 0.001\\
         \hline
    \end{tabular}
    \label{tab:rt_params}
\end{table}
\begin{table}[htbp]
    \centering
    \caption{Path Refinement Threshold Parameters}
    \begin{tabular}{SSSSSSS[table-format=3.0]}
        \hline
        {Case} & {$t_{d}$ (m)} & {$t_{a}$ ($\degree$)} \\
        \hline
        {Noiseless, Ground Truth Normals} & 0.002 & 1.0\\
        {Noisy, Ground Truth Normals} & 0.02 & 1.0\\
        {Noisy, Estimated Normals} & 0.02 & 25.0\\
        \hline
    \end{tabular}
    \label{tab:refine_params}
\end{table}

\subsection{Performance Evaluation}

The simulations were performed on a desktop computer equipped with an AMD Ryzen 9 5900X CPU, NVIDIA GeForce RTX 3080 GPU, and 32GB of RAM. For the simulation parameters, a focus was placed on varying voxel sizes and the number of interactions, emphasizing their influence on performance metrics.

The results of the simulations can be found in Table \ref{table:rt_statistics}. The execution time of pre-processing is less than 200ms for all of the simulations, which is not presented in the aforementioned table. A great increase can be seen in the number of coarse paths when the number of interactions is increased due to the conical rays. As one can expect, processing more interactions results in an increased execution time. It is important to underline that the number of coarse paths only contain a maximum of $\kappa$ duplicate path trajectories. Thus, especially with higher number of interactions, the actual number of coarse paths is higher. Smaller voxel size increases the number of IEs, which increases the resolution of the traversal voxel grid. This results in an increased coarse path tracing execution time, which can be seen in the results. However, the number of found coarse paths decreases with smaller voxel sizes. There are multiple factors that have an effect on the number of found paths. First, the intersection cone angle is determined based on the voxel size. Thus, with a smaller voxel size, the cone apex angle is also smaller. Furthermore, each PCIE intersection considers all of the points contained in the intersection AABB. Because of this, the estimated normal vector may vary between simulations with different voxel sizes. This change of IE contents at a specific spatial location in a scene size may also result in a different label, IE reception point, and even occlusion. Especially, the change of labels may result in varying number of found coarse paths, as only $\kappa$ label combinations (\textit{i.e.} duplicate paths) are allowed for the coarse paths.

\label{ss_performance}
The number of refined paths without a diffraction is equal in all of the simulations. However, when we look at the diffraction simulations, the number of found paths varies between the used voxel sizes. Diffractions are more suspectible to behave differently with different voxel sizes, as the ray is always traced to the IE reception point, meaning that we only have one ray direction from the diffraction edge to a specific IE. Thus, if the IE is close to the diffraction edge, the following ray direction from such IE can vary significantly based on the IE reception point spatial location. This may lead to some paths being missed with certain voxel sizes. Such problematic edge can be seen on the medical supply box on the wall shown in Fig.~\ref{fig:models}. When rays are diffracted from the medical supply box edge towards the wall it lies on, the reflected ray direction may be noticeably different between varying voxel sizes.  

\begin{figure}[t]
    \centering
    \includegraphics[width=0.6\linewidth]{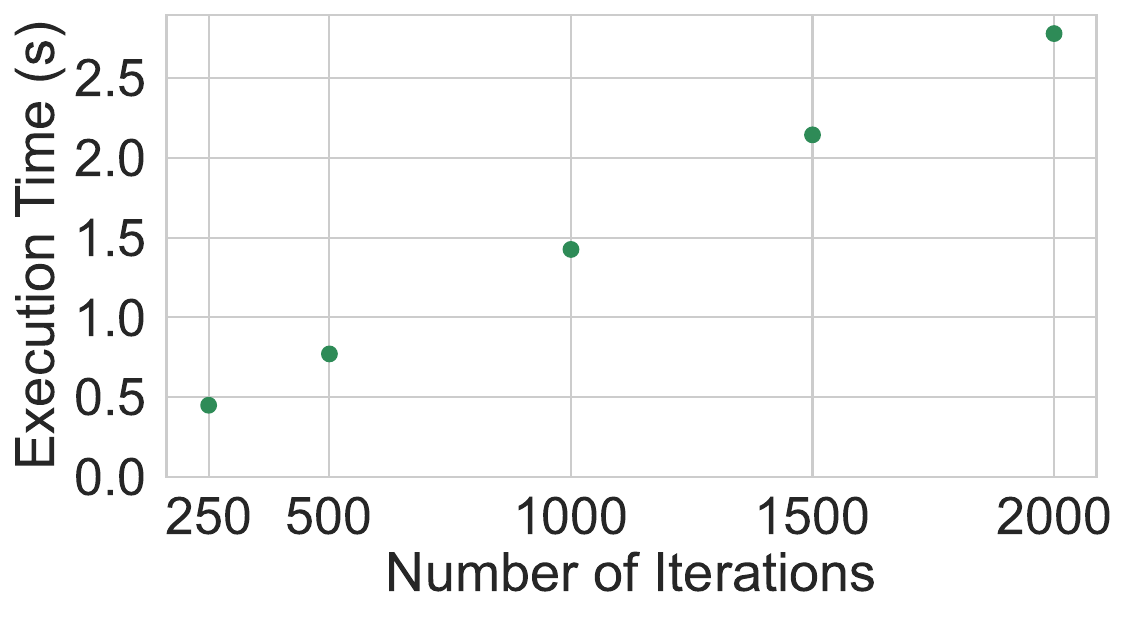}
    \caption{Impact of the number of iterations to path refinement execution time with 105 730 coarse paths. The paths consist of five of fewer reflections.}
    \label{refineIter}
\end{figure}

From the path refinement execution times we can see that they increase when the voxel size is increased. This is due to the intersection program having to evaluate a higher number of points when calculating the intersection point. The performance could be improved by increasing the subvoxel division factor $D_{sv}$ presented in Section \ref{ss_voxelization}. However, increasing $D_{sv}$ has to be done carefully to ensure that the surfaces are still continuous in terms of AABB spatial distribution. This is due to the AABB size minimization procedure mentioned in the aforementioned section. Additionally, the refinement performance was evaluated with 105~730 coarse paths acquired with a limit of 5 reflections and a voxel size of 0.5m$^3$. The results are shown in Fig.~\ref{refineIter}. The number of refined paths was 90 in all simulations except with 250 and 1500 iterations, where 89 paths were found.

The execution time of post-processing is negligible due to the label hashing, which noticeably reduces the number of paths in particular with a high number of interactions. Especially, as only unique label combinations remain after the path refinement phase. The processing only takes some milliseconds in these experiments.

\begin{table*}[t]
    \caption{Number of paths and execution times for different phases of the path determination process with varying voxel sizes}
    \begin{tabular}{SSSSSSS[table-format=3.0]}
        \hline
        {Voxel Size (m$^3$)}  &  {Diffraction Limit} & {Number of Interactions} & {Coarse Paths} & {Refined Paths} & {Coarse Path Tracing (s)} & {Path Refinement (s)}\\
        \hline
            &   & 1 & 7      & 2  & 0.01  & 0.04 \\
            &   & 2 & 193    & 9  & 0.03  & 0.19 \\
        0.4 & 0 & 3 & 1 737  & 25 & 0.07  & 0.42 \\
            &   & 4 & 11 765 & 52 & 0.74  & 0.60 \\
            &   & 5 & 76 725 & 90 & 10.82 & 1.42 \\
        \hline
            &   & 1 & 5       & 2  & 0.01 &  0.06\\
            &   & 2 & 149     & 9  & 0.03 &  0.33\\
        0.5 & 0 & 3 & 1 703   & 25 & 0.06 &  0.60\\
            &   & 4 & 14 085  & 52 & 0.51 &  1.10\\
            &   & 5 & 105 730 & 90 & 7.65 &  2.80\\
        \hline
            &   & 1 & 6       & 2  & 0.01 & 0.10\\
            &   & 2 & 150     & 9  & 0.04 & 0.40\\
        0.6 & 0 & 3 & 1 719   & 25 & 0.10 & 1.33\\
            &   & 4 & 14 793  & 52 & 0.52 & 1.46\\
            &   & 5 & 114 012 & 90 & 5.58 & 4.00\\
        \hline
            &   & 1 & 47      & 5     & 0.01   & 0.04\\
            &   & 2 & 945     & 31    & 0.09   & 0.19\\
        0.4 & 1 & 3 & 7 818   & 141   & 0.64   & 0.66\\
            &   & 4 & 81 448  & 505   & 9.30   & 2.72\\
            &   & 5 & 665 826 & 1 530 & 269.10 & 15.03\\
       \hline
            &   & 1 & 38      & 5     & 0.01   & 0.06\\
            &   & 2 & 731     & 30    & 0.06   & 0.35\\
        0.5 & 1 & 3 & 7 093   & 139   & 0.36   & 0.85\\
            &   & 4 & 82 994  & 492   & 5.13   & 4.87\\
            &   & 5 & 755136  & 1 480 & 150.02 & 26.35\\
        \hline
            &   & 1 & 42      & 5      & 0.01   & 0.12\\
            &   & 2 & 856     & 31     & 0.10   & 0.49\\
        0.6 & 1 & 3 & 8 424   & 145    & 0.43   & 1.54\\
            &   & 4 & 101 303 & 525    & 4.80   & 7.81\\
            &   & 5 & 920 410 & 1 645  & 129.50 & 44.65\\
        \hline
    \end{tabular}
    \label{table:rt_statistics}
\end{table*}

\subsection{Comparison with Wireless Insite}

Wireless Insite is a commercial RT tool which has been validated against measurements in many scenarios, for example, in \cite{mededjovic2012wireless, suga2023indoor, wu2016intra, liu2019path}. In the experiments we compare against the X3D RT model, which generates initial paths by RL with a given ray spacing. The paths are then corrected to exact paths similarly to the image method. Thus, it produces comparable results to our implementation and operates as a baseline for validating our simulated paths. The Wireless Insite simulations were performed with the triangle mesh model presented in Fig.~\ref{fig:models}. For the purpose of comparison, our simulator utilizes the synthetic point cloud version of the triangle mesh model. For each one of the simulated paths, we attempted to find a matching trajectory from the baseline paths. A path was considered to be equal if the angle between the respective rays of the path trajectories were below one degree and if the chain of interaction types was the same. A more detailed comparison is conducted in contrast to our prior work \cite{vaara2024ray}.

\subsubsection{Reflections}

Reflections paths were computed using the limit of 5 interactions. Visual comparison of the simulated paths are provided in Fig.~\ref{totalrefldiff}. Our simulator found 90 paths, while Wireless Insite found 83. All of the baseline paths were found by our simulator. Some of these additional paths are not accurate representations of specular reflections due to the normal vector approximation near an edge.

\subsubsection{Diffractions}

Wireless Insite supports interior edge diffractions, which we do not consider for the reasons explained in Section \ref{ext_edge}. Thus, we removed the baseline paths that contained an interior edge diffraction to enable comparison against our paths.

\begin{figure*}[t]
    \includegraphics[width=1.0\textwidth]{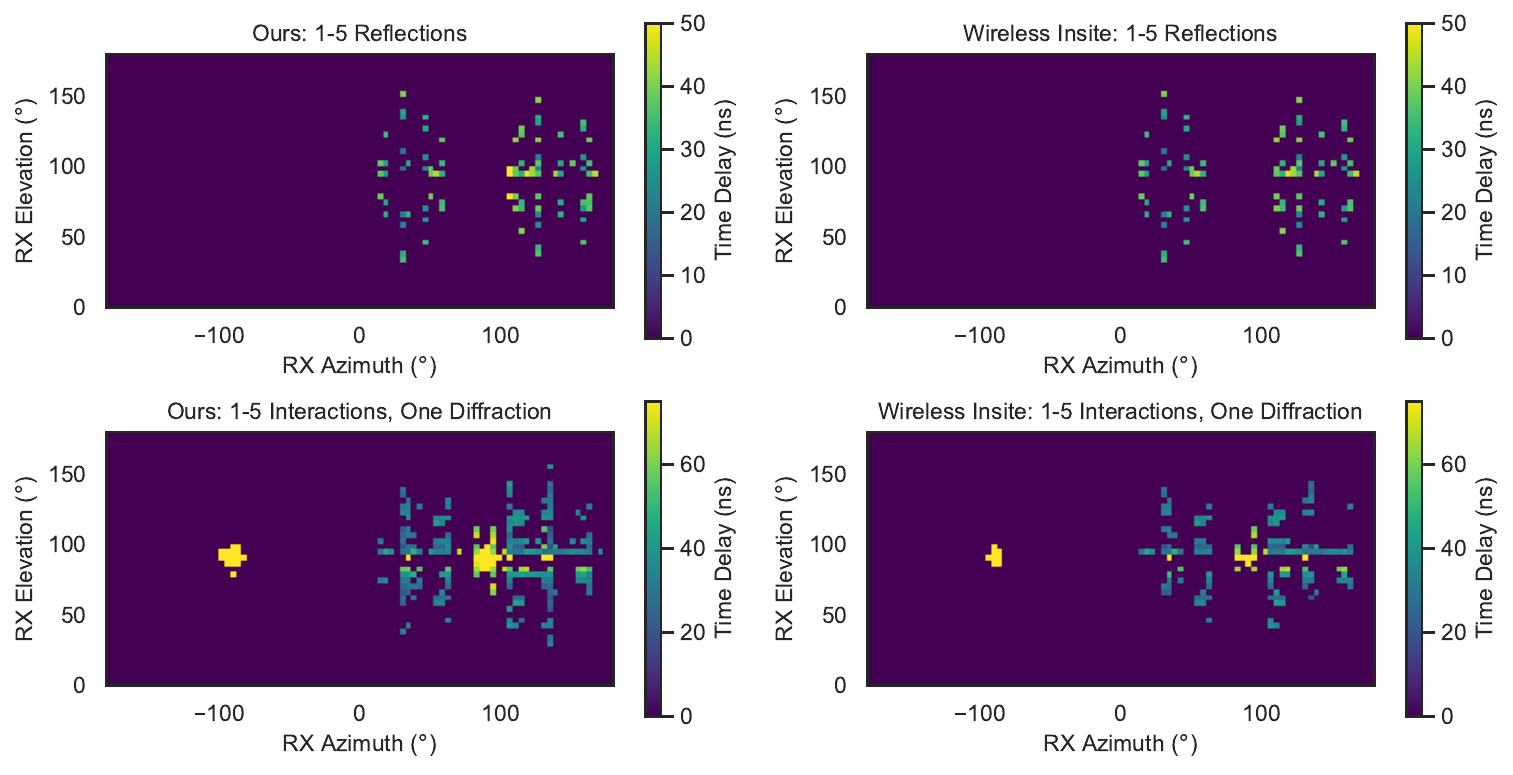}
    \label{totalrefldiff}
    \caption{Paths with 1--5 interactions. For clearer visualization, the heatmap time delays are clamped to 50 and 75 nanoseconds for the upper and bottom row figures, respectively. }
\end{figure*}

The simulations were performed with an interaction limit of 5 with a maximum number of one diffraction. Visual comparison is shown in Fig.~\ref{totalrefldiff}. Our simulator found 1390 paths containing a diffraction, while Wireless Insite found 571. Around 84\% of the baseline paths that contain a diffraction were found by our simulator. As explained in Section \ref{ss_performance}, our simulator may not find all diffracted paths if there are surfaces present very close to the diffraction edge. Similar to the simulations without diffractions, the unique paths found by our implementation may contain inaccurate paths due to the normal vector approximation near edges.

\subsection{Robustness to the Noise of the Synthetic Model}

To simulate reconstructed point clouds, we added Gaussian noise with a standard deviation of 5--20mm to the synthetic point cloud along the ground truth normal vectors to study the impact on the generated path quality. The synthetic point cloud is shown in Fig.~\ref{fig:models}. In addition, we performed experiments by approximating the normal vectors of the noisy point cloud variations. The simulations were performed with the maximum number of reflections limited to 5 and with a voxel size of 0.5m$^3$. The parameters of noisy point cloud simulations can be found in Table \ref{tab:rt_params} and \ref{tab:refine_params}.

\begin{figure*}[htbp]
\centering
\subfloat[Path tracing and refinement performed with ground truth normals.]
{
    \includegraphics[width=0.3\textwidth]{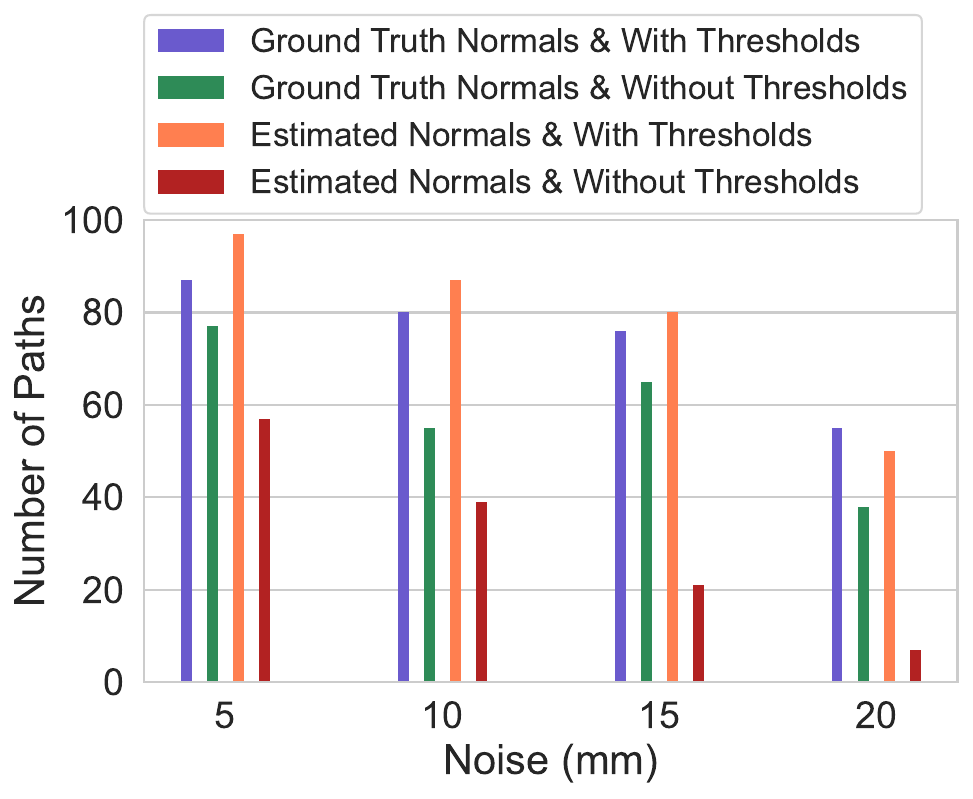}
    \label{fig:numPaths}
}\hspace{0.01\textwidth}
\subfloat[Percentage of noiseless point cloud simulation paths found by the noisy point cloud simulations.]
{
    \includegraphics[width=0.3\textwidth]{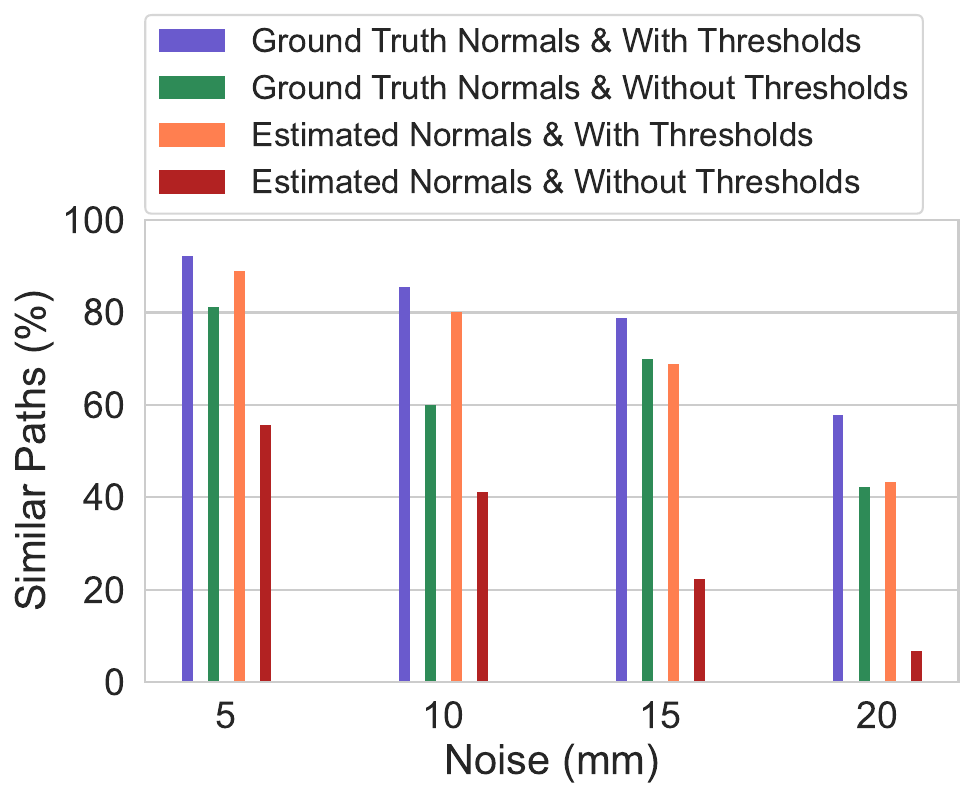}
    \label{fig:matchVct}
}\hspace{0.01\textwidth}
\subfloat[Percentage of baseline paths found by the noisy point cloud simulations.]
{
    \includegraphics[width=0.3\textwidth]{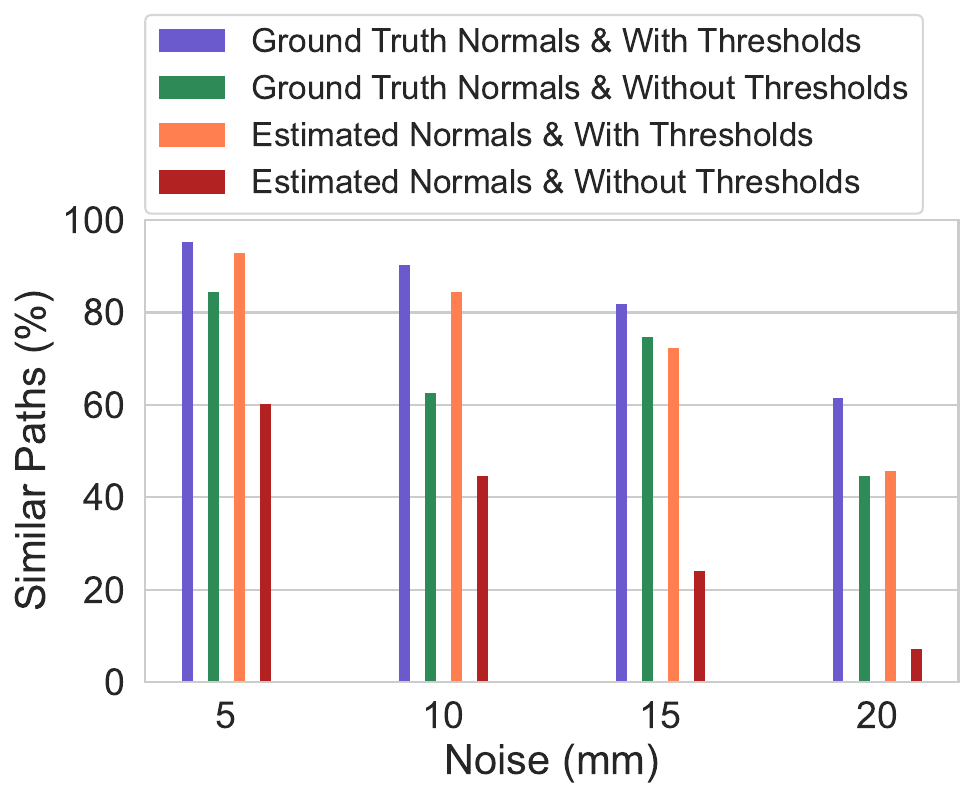}
    \label{fig:matchWi}
}
\caption{Experiments with noise along ground truth normals vectors. The millimeter values represent the standard deviation. Simulations were performed with the number of reflections limited to five.}
\label{noiseExperiments}
\end{figure*}

\begin{figure*}[htbp]
\centering
\subfloat[Number of refined paths.]
{
    \includegraphics[width=0.3\textwidth]{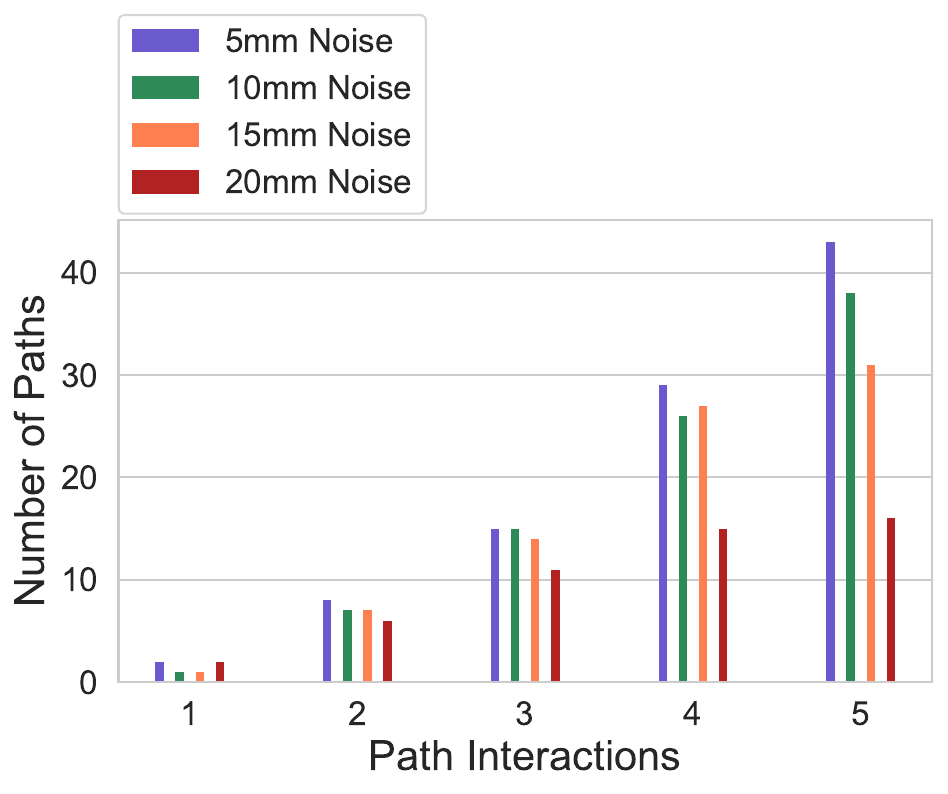}
    \label{fig:iaNumPaths}
}\hspace{0.01\textwidth}
\subfloat[Percentage of noiseless point cloud simulation paths found by the noisy point cloud simulations.]
{
    \includegraphics[width=0.3\textwidth]{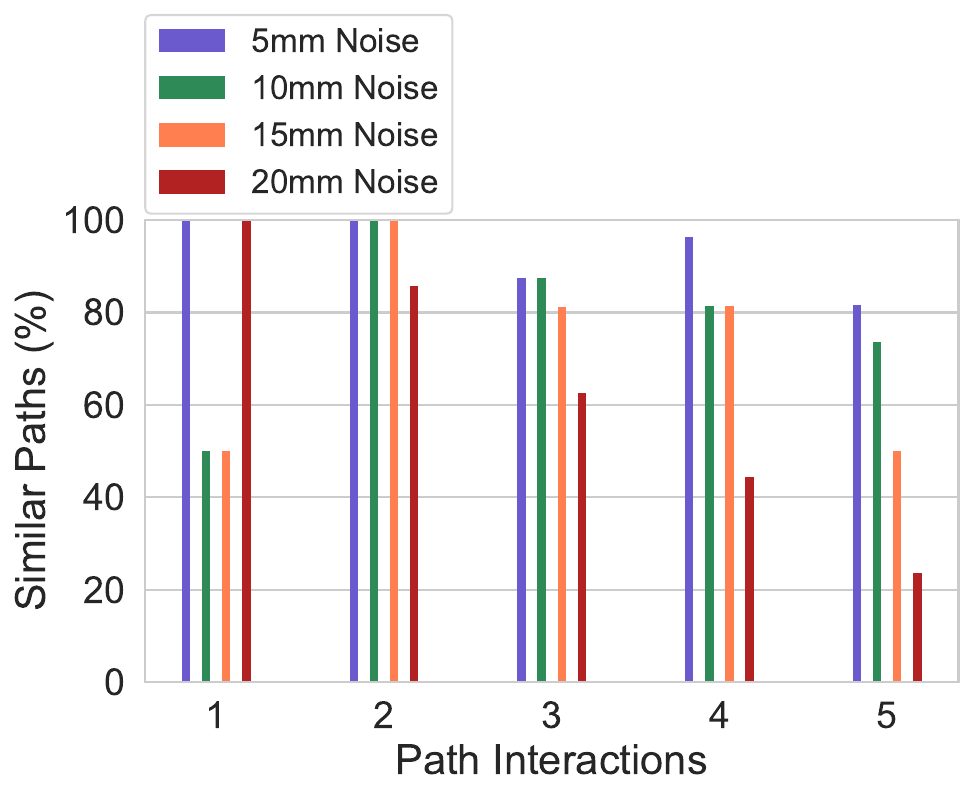}
    \label{fig:vctNoisePercent}
}\hspace{0.01\textwidth}
\subfloat[Percentage of baseline paths found by the noisy point cloud simulations.]
{
    \includegraphics[width=0.3\textwidth]{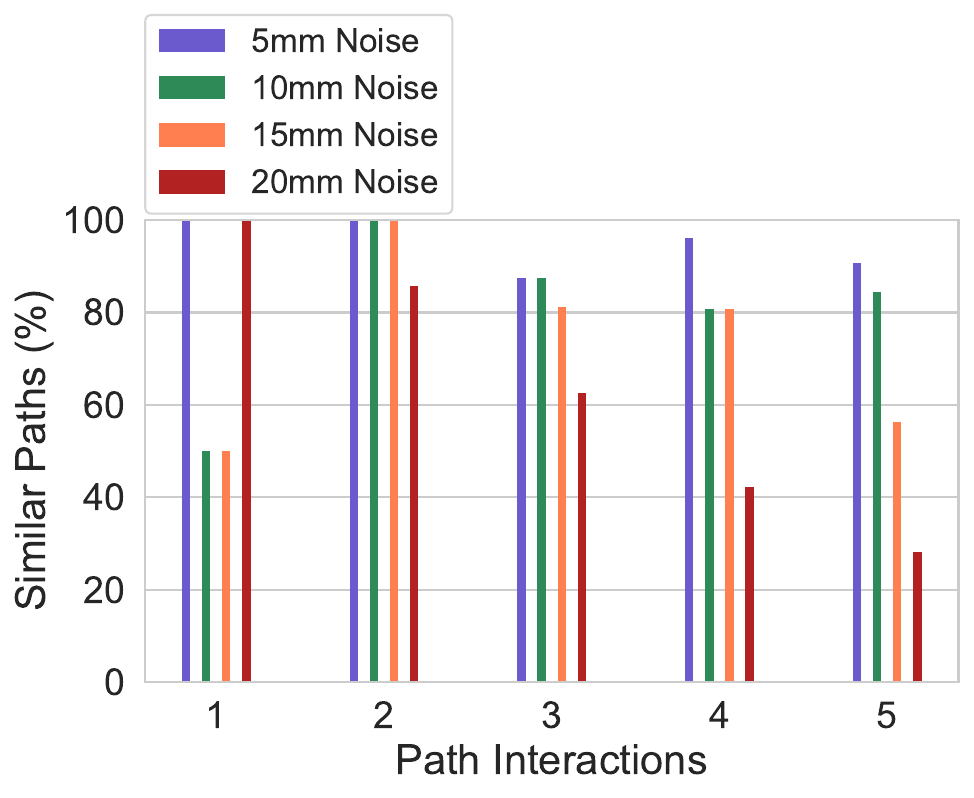}
    \label{fig:wiNoisePercent}
}
\caption{Breakdown of noisy point cloud simulations with thresholds and estimated normal vectors.}
\label{noiseIaPaths}
\end{figure*}

The results of the noisy point cloud experiments are shown in Fig.~\ref{noiseExperiments} and \ref{noiseIaPaths}. The similarity of paths was deemed by comparing the ray directions along a path with a maximum allowed difference of 10 degrees for each ray. Such large angle was picked due to the short distance between the interactions points and the spatial differences caused by the noise. From the results we can see that the effect of noise is small with ground truth normal vectors. Even with small amount of noise, it can be seen that some paths are always missing from the noisy model simulations. One of the factors for this is the larger sample radius $r_{s}$ used in the noisy simulations, which causes fluctuations quicker in the estimated normal vector when the intersection point approaches an edge during the refinement process. The noisy simulation results with ground truth normal vectors are the most accurate when compared to the synthetic model simulations. This underlines the importance of accurate normal vector approximation for point clouds that are utilized by RT-based algorithms. The number of paths without thresholds $t_d$ and $t_a$ is much smaller compared to simulations with thresholds. Especially with 20mm of noise and estimated normal vectors, the simulation only finds 7\% of the baseline paths when not using the thresholds. The same simulation with the thresholds found about 45\% of the baseline paths.

In Fig.~\ref{noiseIaPaths}, the simulated paths with thresholds are shown in terms of the number of interactions. Paths consisting of one interaction were lost in some of the simulations. This is due to the variation caused by noise in the normal vector approximation of the initial interaction point and the spatial variation of IE reception point. In addition, as explained in Section \ref{voxel_cone_tracing}, the discretization may cause degraded accuracy with large voxel size, especially with a low number of interactions. The percentage of found paths in comparison to the synthetic model simulations can generally be seen decreasing when the number of interactions is increased. This is due to the spatial changes in the environment caused by the noise, which results in some of the paths being no longer valid. The noisiness also increases the error in the approximated normal vectors. Due to this effect, some of the paths may not be able to converge during the refinement process with the noisy point clouds.

\subsection{Path Comparison with Channel Measurements}

The channel measurements were performed in \cite{kokkoniemi2022initial}, thus, we only provide a short description of the setup. The measurements were performed at the D band, which covers the frequencies from 110 to 170 GHz. The channel sounding measurement system consist of vector network analyzer (VNA), VNA extension modules, and custom 3D rotation stages. The used antennas are horn antennas with a half power beam width of 10 and 9 degrees in azimuth and elevation, respectively. We compare against the NLOS measurement, which covers an area of 90 degrees in azimuth and 85 degrees in elevation at the RX side. Each measured azimuth and elevation was scanned for a total of 66.7ns with 4001 samples per scan (0.017ns per sample). To compare our paths to the measurements, an aggregated impulse response was computed by picking the maximum sample value of each measured impulse response at a given RX azimuth and elevation angle.

\begin{figure}[htbp]
    \centering
    \includegraphics[width=1.0\linewidth]{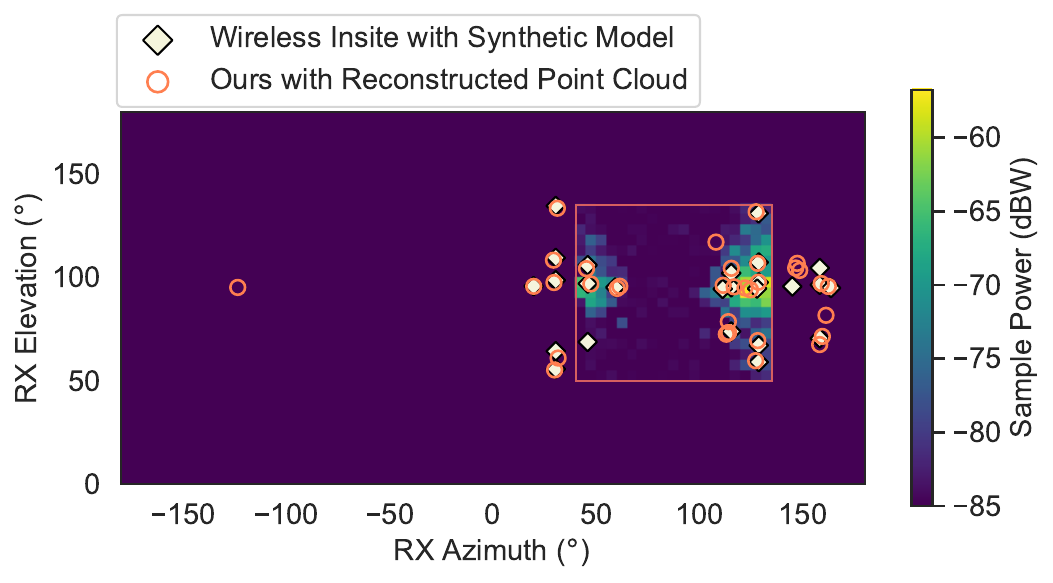}
    \caption{Aggregated impulse response. The red rectangle represents the measured azimuth and elevation angles at RX. The ray traced paths consist of three or fewer reflections.}
    \label{fig:meascomp}
\end{figure}

The simulation was performed with the reconstructed point cloud presented in Fig.~\ref{fig:models}, which contains noise and estimated normal vectors. The simulation specific parameters for such point cloud can found in Table~\ref{tab:rt_params} and Table~\ref{tab:refine_params}. The number of interactions was restricted to three reflections. The label generation in the case of reconstructed point cloud resulted in some walls being a mix of multiple labels, which can be seen as small clusters in the results presented in Fig.~\ref{fig:meascomp}. In addition, we simulated the same scenario with the synthetic model (Fig.~\ref{fig:models}) using Wireless Insite. The simulations were not restricted to the measurement azimuth and elevation angles, which is why paths outside of the measurement angles can be seen in the figure. The results show similarity between our paths and the paths found by Wireless Insite with the synthetic model, though some differences are expected due to more detailed geometry, noise, and estimated normal vectors present in the reconstructed point cloud. Both of the simulated path sets form an analogous pattern to the aggregated impulse response within the measured azimuth and elevation angles, which further validates the ray traced path trajectories. One outlier path can be seen between -150 and -100 degrees in azimuth, which is caused by a reflection from a recessed area near a door. Such details are not present in the synthetic model.

\section{Discussion}
\label{section_discussion}

The results show that our simulator is capable of finding a significant majority of the baseline paths computed by a commercial ray tracer. In addition, we demonstrated that our simulator is capable of finding realistic paths with noisy point clouds that can be constructed from sensor data. This validates the applicability of computer vision-based models for radio channel characterization.

From the experiments it can be seen that the quality of the paths can be noticeably improved if normal vectors are estimated as accurately as possible. In addition, equation~(\ref{nx}) causes rounding of normals around edges due to considering all points contained in the intersected AABB. To counter this, the points within IEs could be divided into distinct AABBs based on the label. Another option could be estimating the weighted curvature in a similar manner to equation~(\ref{nx}), as curvature for each point is acquired when estimating the normal vectors. Paths surpassing a certain curvature threshold at an interaction point could then be discarded. For RT applications, an edge aware normal vector estimation method would be the most practical solution to improve the quality and number of paths. Such method could also incorporate the automatic detection of diffraction edges from the point cloud.

The plane fitting for labeling works sufficiently in the corridor scene. However, in the reconstructed point cloud some walls contain multiple labels due to the distance threshold for plane fitting, which had to be set as the point cloud contains noise. If the scene geometry would be more complex, a plane fitting algorithm for labeling would not be a valid approach.

The number of paths containing diffractions could be increased by additionally evaluating the exact intersection point for each diffracted ray during coarse path tracing or by decreasing the voxel size. This would increase the computational complexity. However, as the mm-wave band is the primary frequency range intended to be used in our simulator, the overhead could be reduced by casting diffracted rays only near the reflection boundary and especially near the shadow boundary, where the scattering is strongest, as noted in \cite{jacob2012diffraction}.

Point clouds are an interesting application for RT, however, triangles are the most used primitives to represent a scene. Although time consuming, accurate models with such primitives can be crafted. Our implementation could be extended to support triangle-based meshes, which would result in a general purpose RT solution for radio channel characterization. In addition, such a simulator should support the remaining propagation effects that are not implemented, namely diffuse scattering and penetration.

\section{Conclusion}
\label{section_conclusion}

We have presented a ray launching algorithm that directly utilizes point clouds as the geometry representation. Utilizing conical rays and environment discretization, 
the method first computes coarse paths, which are then refined to exact paths using a gradient descent based approach.

In experiments, computing performance was assessed by simulations with varying parameters and results demonstrate the efficiency of the method. The focus in validation was placed on assessing the path trajectories, which are the cornerstone of accurate ray tracing-based simulations. Paths computed by a commercial ray tracer, which acted as the baseline, exhibited a significant similarity with the paths computed by the proposed method. In addition, the robustness to noise was assessed by random displacement of points. It was observed that the proposed method is capable of adapting to the introduced noise. Furthermore, it was shown that the method can be utilized with point clouds constructed from RGB-D images. This was demonstrated by comparing the obtained paths to an aggregated impulse response acquired from channel measurements. The findings conclude that the paths exhibit an analogous pattern to the aggregated impulse response, while also illustrating similarity with the baseline path trajectories.

The work provides a solid foundation for applying computer vision-based point clouds to radio channel characterization. In the future, there is a need to research point cloud segmentation to enable support for duplicate path removal in the case of complex geometries. Additionally, automatic diffraction edge detection and enhanced normal vector estimation should be incorporated to improve the accuracy of path trajectories. 
Also, support for propagation mechanisms such as diffuse scattering and penetration will be considered.

\section*{Acknowledgment}
This research was supported by the Research Council of Finland (former Academy of Finland) 6G Flagship Programme (Grant Number: 346208), and the Horizon Europe CONVERGE project (Grant 101094831).

\bibliographystyle{IEEEtran}
\bibliography{citations}

\end{document}